\documentclass[a4paper, amsfonts, amssymb, amsmath, reprint, showkeys, nofootinbib, onecolumn,twoside]{revtex4-1}
\usepackage[english]{babel}
\usepackage[utf8]{inputenc}
\usepackage[colorinlistoftodos, color=green!40, prependcaption]{todonotes}
\usepackage{amsthm}
\usepackage{mathtools}
\usepackage{physics}
\usepackage{xcolor}
\usepackage{graphicx}
\usepackage[left=23mm,right=13mm,top=35mm,columnsep=15pt]{geometry} 
\usepackage{adjustbox}
\usepackage{placeins}
\usepackage[T1]{fontenc}
\usepackage{lipsum}
\usepackage{csquotes}

\usepackage[pdftex, pdftitle={Article}, pdfauthor={Author}]{hyperref} % For hyperlinks in the PDF
\begin{document}
\title{50 Years of Horndeski Gravity: Past, Present and Future}

\author{Gregory W. Horndeski}
    \email{ghorndeski@uwaterloo.ca or horndeskimath@gmail.com}% Your name
    \affiliation{Applied Math Department, University of Waterloo, 200 University Avenue, Waterloo, N2L 3G1, Ontario, Canada}

    \author{Alessandra Silvestri}
    \email{silvestri@lorentz.leidenuniv.nl}% Your name
    \affiliation{Institute Lorentz, Leiden University, PO Box 9506, Leiden 2300 RA, The Netherlands}

\date{\today} % Leave empty to omit a date

\begin{abstract}
An essay on Horndeski gravity, how it was formulated in the early 1970s and how it was 're-discovered' and widely adopted by Cosmologists more than thirty years later.
\end{abstract}

\maketitle
\section{Introduction}
In September, 2023, Prof. Andreas Wipf, the Editor in Chief of the International Journal of Theoretical Physics, contacted me (Gregory Horndeski) concerning my paper entitled "Second-Order Scalar-Tensor Field Equations in a Four-Dimensional Space." That paper originally appeared in his Journal back in 1974, and he thought it would be appropriate for me to write an essay, to appear in the International Journal of Theoretical Physics in 2024, to celebrate the 50th anniversary of that paper's publication. At first I was a little apprehensive about accepting his proposal because he wanted the essay to cover how the theory came into being and how it has been used. Well I certainly was familiar with the theory's inception, but I really was not qualified to discuss how it was being used.  To allay my concerns Prof. Wipf said that he would not mind if the essay had a co-author who could explain the theories value to the Physics community.  Well given that I would be permitted to have a colleague to do the hard work, I agreed to participate in this essay project.  My next task was to find a co-author.  To that end I contacted a colleague of mine at the University of Waterloo, Prof. Ghazal Geshnizjani, who has worked with my scalar-tensor equations, to see if she could provide me with a list of names of physicists who might be able to assist me in my endeavor.  She provided me with the names of a handful of physicists, all of whom were familiar to me from what little I knew about applications of my equations. At the top of the list was Prof. Alessandra Silvestri, with whom I had prior contact concerning my work as an artist.  So I asked her if she would be interested in co-authoring an essay with me and she said yes. 

This joint project between Prof. Silvestri and me will consist basically of two parts.  In the next Section I shall discuss how I developed my scalar-tensor equations.  This will then be followed by a section, with several parts, in which Prof. Silvestri elucidates upon the cosmological implications of what has become known as "Horndeski Gravity, " or "Horndeski Scalar Theory," or "Horndeski Theory." The essay will end with us commenting on the possible future for scalar-tensor theories. 

Since this is an essay, and not a survey article, we shall certainly not be trying to discuss all possible ways in which my equations have been employed.  We just want to give the reader a feel for what has been going on during the past 50 years. Consequently we shall not be able to mention everyone who has made a valuable contribution to the theory's development. Those whose work has not been cited in this essay should not be chagrined, since you were not left out because of malice, but because there are just so many of you.

\section{Horndeski Gravity: the formulation}
I arrived at the University of Waterloo in June, 1970, when I was 22 years old and had just completed my undergraduate work, receiving a B.Sc. in Engineering Physics from Washington University of St.Louis. Prof. Lovelock was assigned to be my supervisor, and at our first meeting it was clear that I needed to learn tensor calculus.  So he asked me to work through the first two chapters of Eisenhart's book "Riemannian Geometry"~\cite{Eisenhart:1964}.  I did that in about three weeks and afterwards visited Prof. Lovelock to find out what comes next.  He promptly went up to the blackboard in his office and wrote down the 5x5 generalised Kronecker delta, which is the determinant of a 5x5 matrix whose entries are Kronecker deltas, and asked me what does that equal in a space of four-dimensions. I had never seen that creature before in my life and really had no idea of what it might equal.  But then I quickly remembered something I had learned in an algebra course I had taken about eight months earlier. That course was taught by a post-doc from the University of Chicago, and he told me, Greg, if someone ever writes something down on a blackboard and asks you what it equals and you are not sure, you say 0 if you are in a vector space, and the identity if you are in a group.  How could he have been so prescient! So I told Prof. Lovelock that the tensor in question equaled 0.  He was slightly taken aback by my reasonably quick response, and then went on to explain why it equaled 0, as if he needed to convince me that he knew why it vanished. He then went on to discuss how the generalised Kronecker delta could be used to prove various things, such as the Cayley-Hamilton Theorem.  Little did I know at that time how important the generalised Kronecker delta would be in my later studies.

The next thing Prof.Lovelock wanted me to learn was some general relativity.  To that end I was required to read and work through most of Adler, Bazin and Schiffer's (1965) book "An Introduction to General Relativity"~\cite{Adler:1965}.  Once that was done I was ready to begin reading Prof. Rund's papers on tensorial concomitants~\cite{Rund:1964,Rund:1966}, and Prof. Lovelock's papers dealing with the uniqueness of the Einstein field equations. While doing that I am sure that Prof. Lovelock was looking for a topic that I could write my Masters thesis on.  That is when he came upon this paper of Bergmann~\cite{Bergmann:1968ve} dealing with the Brans-Dicke Theory~\cite{Brans:1961sx}, and scalar-tensor field equations. The equations of such theories were required to be derivable from a variational principle in which locally the field variables were a scalar function $\phi$ and the components of a Lorentzian metric tensor, $g_{\alpha\beta}$. Bergmann claimed that in a four-dimensional space the most general second-order scalar-tensor field equations derivable from a Lagrangian of the form 
\begin{equation} 
\label{gen_lagrangian}
    L=L(g_{\rm \alpha\beta};g_{\rm\alpha\beta,\gamma};g_{\rm \alpha\beta,\gamma\delta};\varphi;\varphi_{,\rm \gamma})  
\end{equation} 
can be obtained from the Lagrangian
\begin{equation}
\label{L_B}
    L_{\rm B}:=\sqrt{-g}\left(f_1R+f_2X+f_3\right)
\end{equation}
where $f_1, f_2, f_3$ are arbitrary scalar functions of $\varphi$, a comma denotes a partial derivative with respect to local coordinates, $g:=det(g_{\alpha\beta})$, $R:=g^{\alpha\beta}R_{\alpha\beta}$, $R_{\alpha\beta}:=R_{\alpha\,\,\,\beta\gamma}^{\,\,\,\gamma}$, $R_{\alpha\,\,\,\beta\gamma}^{\,\,\,\iota}:=\Gamma^\iota_{\alpha\beta,\gamma}-\Gamma^{\iota}_{\alpha\gamma,\beta}+\Gamma^\mu_{\alpha\beta}\Gamma^\iota_{\mu\gamma}-\Gamma^\mu_{\alpha\gamma}\Gamma^\iota_{\mu\beta}$,  $\Gamma^\iota_{\alpha\beta}:=1/2\,g^{\iota\mu}(g_{\alpha\mu,\beta}+g_{\mu\beta,\alpha}-g_{\alpha\beta,\mu}), g^{\iota\mu}$ is the matrix inverse of $g_{\iota\mu}$, and $X:= -1/2 \,g^{\alpha\beta}\phi_{,\alpha}\phi_{,\beta}$.  Clearly this result is erroneous since we can replace $f_2X+f_3$ in $L_B$ by a more general arbitrary function $F=F(\varphi,X)$ and still get second-order field equations.  In my Master's thesis (1971) I developed, using the aforementioned  work of Rund, the machinery necessary to formally study Lagrangians of the form presented in Eq.\ref{gen_lagrangian}. However the bulk of my thesis dealt with the Brans- Dicke theory, which is derivable from the Lagrangian 

\begin{equation}
\label{L_BD}
    L_{\rm BD}:=\sqrt{-g}\left(\varphi R+2\omega X/\varphi\right)
\end{equation}
with $\omega$ being a constant.  After finishing that thesis in early 1971, Prof. Lovelock and I turned our attention to proving that in a Lorentzian space of four-dimensions the most general Lagrangian of the form given in Eq.\ref{gen_lagrangian} that yields second-order scalar-tensor field equations is given by
\begin{eqnarray}
\label{L_ST_v1}
    L=&&\sqrt{-g}\left\{\beta_1\left[R^2-4R_{\rm \kappa\lambda}R^{\rm\kappa\lambda}+R_{\rm \kappa\lambda\mu\nu}R^{\rm \kappa\lambda\mu\nu}\right]+\beta_2G^{\rm \mu\nu}\varphi_{\rm ,\mu}\varphi_{\rm, \nu}+\beta_3R+\eta\right\}\nonumber\\
    \nonumber\\
    &&+\alpha \epsilon^{\rm \kappa\lambda\mu\nu}
    R_{\,\,\,\,\,\,\rm \kappa\lambda}^{\rm \theta\iota}R_{\rm\theta\iota\mu\nu}  
\end{eqnarray}
where $\alpha$ is a real constant, $\beta_1,\ \beta_2,\   
\beta_3$ are functions of $\phi$ and   $\eta = \eta 
 (\phi\ ,X)$. 
 
Our proof of the validity of Eq.\ref{L_ST_v1} appeared in Horndeski and Lovelock~\cite{Horndeski:1972} (which was a special issue of the Journal of the Tensor Society celebrating the 70th birthday of the renowned Japanese Mathematician A. Kawaguchi) and was a non-trivial generalization of Lovelock's papers~\cite{Lovelock:1969, Lovelock:1970}. There he established (among other things) that:

(i) In a space of four-dimensions the most general Lagrangian of the form $L = L(g_{\alpha\beta}; g_{\alpha\beta,\mu}; g_{\alpha\beta,\mu\nu})$ which yields second-order field equations is
\begin{equation}
\label{L_L}
    L_{\rm L}:=\sqrt{-g}\left\{\lambda R+2\Lambda+\beta\left[R^2-4R_{\rm \kappa\lambda}R^{\rm\kappa\lambda}+R_{\rm \kappa\lambda\mu\nu}R^{\rm \kappa\lambda\mu\nu}\right]\right\}+\alpha\epsilon^{\rm\kappa\lambda\mu\nu}R_{\,\,\,\,\,\,\rm\kappa\lambda}^{\rm\theta\iota}R_{\rm\theta\iota\mu\nu}
\end{equation} 

where $\alpha$, $\beta$, $\lambda$ and $\Lambda$ are constants; and

(ii) In a space of four-dimensions the most general symmetric, contravariant tensor density $A^{\alpha\beta} = A^{\alpha\beta}(g_{\gamma\theta}; g_{\gamma\theta,\mu}; g_{\gamma\theta,\mu\nu})$ which is divergence-free, is given by

\begin{equation} 
\label{tensor_density}
    A^{\rm\alpha\beta}=\sqrt{-g}\left\{-\lambda G^{\rm \alpha\beta}+\Lambda g^{\rm \alpha\beta}\right\}\,,
\end{equation}

In fact $A^{\alpha\beta}$ is the variational derivative of $L_L$.

In passing I would like to point out that in Eq.\ref{L_L} the $\beta$ Lagrangian is proportional to the Gauss-Bonnet Lagrangian, and the $\alpha$ Lagrangian is proportional to the Pontryjagin Lagrangian.  Both of these Lagrangians have identically vanishing Euler-Lagrange tensors.

In view of my work in my Master's thesis and the paper I wrote with Prof. Lovelock; Prof. Rund (who was the chairman of the Applied Math Department at the University of Waterloo at the time), and Prof. Lovelock decided that I was qualified to enter the Ph.D. program in 1971.  It was also quite clear what the task of my thesis would be; $viz.,$ to construct in a space of four-dimensions the most general second-order scalar-tensor field equations that can be obtained from a variational principle using a Lagrangian of arbitrary differential order in the derivatives of $\phi$ and $g_{\alpha\beta}$.  The approach I took in solving this problem was modeled on Prof. Lovelock's technique to establish (ii) above, and goes as follows.

     If $L$ is a Lagrangian of the form
\begin{equation}
\label{higher_gen_Lagrangian}
    L=L(g_{\rm \alpha\beta};g_{\rm \alpha\beta,\gamma};g_{\rm \alpha\beta,\gamma\delta};\cdots;\varphi;\varphi_{,\rm\gamma};\cdots)
\end{equation}
(where the derivatives of $g_{\rm\alpha\beta}$ and $\varphi$ stop at some point) then the Euler-Lagrange tensor densities
\begin{equation}
\label{EL_1}
E^{\rm \alpha\beta}(L)\equiv\frac{\delta L}{\delta g_{\rm \alpha\beta}}:=\frac{\partial L}{\partial g_{\rm\alpha\beta}}-\frac{d\,\,}{dx^{\rm \gamma}}\frac{\partial L}{\partial  g_{\rm \alpha\beta,\gamma}}+\cdots 
\end{equation} 
and
\begin{equation}
\label{EL_2}
    E(L)\equiv\frac{\delta L}{\delta\varphi}:=\frac{\partial L}{\partial\varphi}-\frac{d\,\,}{dx^{\rm \gamma}}\frac{\partial L}{\partial\varphi_{,\rm \gamma}}+\cdots
\end{equation}
are related by 
\begin{equation}
\label{EL_rel}
    E^{\rm \alpha\beta}(L)_{|_\beta}=\frac{1}{2}\varphi^{\rm \alpha}E(L)\
\end{equation}
where the vertical bar, $_|$ denotes covariant differentiation.  (The proof of Eq.\ref{EL_rel}  is given in~\cite{Horndeski:1973}, but those who use infinitesimal variations can derive it from the invariance of the Lagrangian under the coordinate transformation that replaces $x^\iota$, by   $x^\iota + \xi^\iota.$)  Eq.\ref{EL_rel} shows us that if the field tensors are of second-order then the divergence of $E^{\alpha\beta}(L)$  must  also be of second-order, and not of third order.  This leads us to consider the following problem:  In a space of four-dimensions construct all second-order symmetric (0,2) tensor densities

\begin{equation}
\label{tensor_density_2}
    A^{\rm \alpha\beta}=A^{\rm \alpha\beta}(g_{\rm \gamma\delta};g_{\rm \gamma\delta,\mu};g_{\rm \gamma\delta,\mu\nu};\varphi;\varphi_{,\rm \mu};\varphi_{,\rm \mu\nu})
\end{equation} 
which are such that there exists a second-order scalar-tensor concomitant A with the property that

\begin{equation}
\label{conc_tensor}
    A^{\rm \alpha\beta}_{\,\,\,\,\,\,|\rm \beta}=\varphi^{\rm \alpha}A\,.
\end{equation}

So not only are we requiring the divergence of $A^{\alpha\beta}$ to be of second-order, but it must be parallel to $\phi^\alpha$.  

By generalizing the techniques used in~\cite{Lovelock:1970} I was able to construct, using numerous generalized Kronecker deltas, all of the tensorial concomitants $A^{\alpha\beta}$ and A satisfying Eq.\ref{tensor_density_2} and Eq.\ref{conc_tensor}. At that point we know that all of the field tensors for a second-order scalar-tensor theory will be contained somewhere in $A^{\alpha\beta}$ and A, but we are not sure if there exists a Lagrangian that will yield $A^{\alpha\beta}$  itself as its Euler-Lagrange tensor.  Well, when Lovelock  was confronted with a similar problem he just said, why not try the scalar density $g_{\alpha\beta}A^{\alpha\beta}$ as a possible Lagrangian.  So that is what I did and with some effort it worked!  (Why it works will be explained by me in a future publication.)  In this way I was able to come up with a Lagrangian and an algorithm that would show how to go from any $A^{\alpha\beta}$ and A that satisfies Eq.\ref{tensor_density_2} and Eq.\ref{conc_tensor}, to a Lagrangian L, which was such that $E^{\alpha\beta}(L) = A^{\alpha\beta}$. This result first appeared in print in~\cite{Horndeski:1974wa}.  In terms of the notation currently used (with $\nabla$ now denoting covariant differentiation), my original Lagrangian yielding the most general second-order scalar-tensor field equations in a four-dimensional space is expressible as
\begin{eqnarray}
    \label{Horndeski_original}
\mathcal{L}_H&=&\sqrt{-g}\Big\{\delta^{\alpha\beta\gamma}_{\mu\nu\sigma}\Big[\kappa_1\nabla^\mu\nabla_\alpha\phi R_{
    \beta\gamma}^{\,\,\,\,\,\,\nu\sigma}+\frac{2}{3}\kappa_{1X}\nabla^\mu\nabla_\alpha\phi\nabla^\nu\nabla_\beta\phi\nabla^\sigma\nabla_\gamma\phi+\kappa_3\nabla_\alpha\phi\nabla^\mu\phi R_{
    \beta\gamma}^{\,\,\,\,\,\,\nu\sigma} 
    \nonumber\\ &+&2\kappa_{3X}\nabla_\alpha\phi\nabla^\mu\phi\nabla^\nu\nabla_\beta\phi\nabla^\sigma\nabla_\gamma\phi\Big]+\delta_{\mu\nu}^{\alpha\beta}\Big[(F+2W)R_{   \alpha\beta}^{\,\,\,\,\,\,\mu\nu}+2F_X\nabla^\mu\nabla_\alpha\phi\nabla^\nu\nabla_\beta\phi\nonumber\\
&+&2\kappa_8\nabla_\alpha\phi\nabla^\mu\phi\nabla^\nu\nabla_\beta\phi\Big]-6(F_\phi+2W_\phi-X\kappa_8)\Box\phi+\kappa_9\Big\},
\end{eqnarray}    
where    $\delta^{\alpha_1\alpha_2\dots\alpha_n}_{\mu_1\mu_2\dots\mu_n}=n!\delta^{[\alpha_1}_{\mu_1}\delta^{\alpha_2}_{\mu_2}\dots\delta^{\alpha_n]}_{\mu_n}$ and $\kappa_1, \kappa_3, \kappa_8, \kappa_9$ are arbitrary functions of $\phi$ and $X$. $F=F(\phi,X)$ is determined up to a quadrature by the equation $F_X = 2(\kappa_3 + 2X\kappa_{3X} - \kappa_{1\phi})$, with $W=W(\phi)$ essentially being an integration "constant" that arises when computing F. $\phi$ and $X$ at the subscript of the functions indicate derivation of the latter with respect to the scalar field and its kinetic term, respectively. In \textbf{Section 3} it will be explained how one goes from the above Lagrangian,  to the one currently favored.   
   
Upon comparing the Lagrangian ${\mathcal L}_H$ with the one presented in Eq.\ref{L_ST_v1}, one can't help but wonder why there isn't a term in ${\mathcal L}_{H}$  involving a Lagrangian that is quadratic in the curvature tensor.  The answer is that the quadratic Gauss-Bonnet type Lagrangian can actually be built from the Lagrangians in Eq.\ref{Horndeski_original} (see Appendix G in~\cite{Horndeski:1973}), and the Pontryagin Lagrangian has an identically vanishing Euler-Lagrange tensor, and has been dismissed.

It is interesting to note that if one demands that the field equations derivable from the Lagrangian presented in Eq.\ref{Horndeski_original} are quasi-linear in the second derivatives of the field variables; \emph{i.e.}, the coefficients of $g_{\alpha\beta,\gamma\delta}$ and $\varphi_{,\gamma\delta}$ in the field equations are at most functions of $g_{\alpha\beta}$ and $\phi$, then (as I point out in~\cite{Horndeski:1974wa}) we recover the Bergmann Lagrangian given in Eq.\ref{L_B}, up to a divergence.  It is the quasi-linearity of the second-order 
 scalar-tensor field equations which uniquely characterizes Bergmann's Lagrangian, up to a divergence.

 When Prof. Lovelock and I saw how complex the Lagrangian which yields the most general second-order scalar-tensor field equations in a space of four-dimensions were, we felt that clearly puts the kibosh on scalar-tensor field theories. 
 There were just too many of them, and they are way too complicated.  We wondered who would be crazy enough to work with such equations.  Then crazy showed up!  And is still here today.  It will be the task of my colleague, Prof. Silvestri to  explain why these equations became so useful, after languishing as mere mathematical curiosities for over thirty years, before being rediscovered by Charmousis, \textit{et al}~\cite{Charmousis:2011bf}.

\section{Horndeski Gravity in Cosmology}
The discovery of cosmic acceleration in 1998~\cite{SupernovaSearchTeam:1998fmf,SupernovaCosmologyProject:1998vns} provided a new boost for the  exploration of alternatives to general relativity (GR) in the description of gravity, this time in the cosmological context. In fact,  a phase of accelerated expansion consistent with the observations, within a Universe governed by GR, requires a mysterious source  which permeates space in an everlasting manner. In other words, a component with  pressure equal to minus its energy density, i.e. characterized by an equation of state $w\sim-1$\footnote{The equation of state for a given species is defined as the ratio between the pressure and energy density of that species, i.e. $w\equiv P/\rho$.}. The constituents we are familiar with, i.e. matter in the form of dust and radiation, are  characterised by $w=0$ and  $w=1/3$, respectively and cannot source an accelerated expansion. The vacuum energy of quantum fields is an obvious, and perhaps inescapable, candidate which shall contribute a constant energy density to the budget of the Universe, of the order $\Lambda_{\rm th}\sim 10^{-60}-1\,{\rm M}_P^4$ depending on the ultraviolet cutoff on the theory~\cite{Weinberg:2000yb,Joyce:2014kja}.   Measurements of the expansion rate of the Universe give $\Lambda_{\rm obs}\sim 10^{-120}\,{\rm M}_P^4$ which is $15-30$ orders of magnitude smaller, in mass scale. This would require an unnatural fine tuning of the bare constant in the action, $\Lambda_0$, so to cancel almost perfectly the huge vacuum energy of quantum fields, leaving a very small contribution that matches $\Lambda_{\rm obs}$.  This discrepancy,  known as the \emph{cosmological constant problem}, has prompted the exploration of dynamical scenarios that would alleviate the fine tuning by employing a \emph{dark energy} (DE) field~\cite{Copeland:2006wr},  and of modifications of the laws of gravity (MG) that would provide self-accelerating solutions in absence of any matter~\citep{Clifton:2011jh} and, ideally, degravitate the vacuum energy~\cite{Dvali:2007kt}. 
This activity intensified through the years, leading to an intricate  landscape of candidate models of gravity and DE. We refer the reader to~\citep{Clifton:2011jh} for a comprehensive review of the latter.

While GR has been confirmed to great accuracy  in the laboratory and in the Solar System~\cite{Will:2014kxa}, and, more recently, with the direct detection of gravitational waves~\cite{Abbott:2016blz,TheLIGOScientific:2017qsa} and the imaging of black holes~\cite{EventHorizonTelescope:2019dse}, none of these tests probe gravity on cosmological scales, where it is characterised by the Hubble expansion of the Universe. This is  a completely new regime, corresponding to vastly different lengthscales and curvatures, and the diffuse presence of matter.  We shall approach it with the same level of scrutiny achieved in local tests of gravity. From the observational point of view,  this will become a concrete possibility with the advent of Stage IV Large Scale Structure (LSS) surveys, such as Euclid and the Vera C. Rubin Observatory’s Legacy Survey. In anticipation of this endeavour, on the theoretical side it has become gradually clear that we cannot proceed on a model basis, but we rather need  unifying  frameworks for a fundamentally driven exploration of gravity. 
In fact, a valuable lesson learned from local tests of gravity, is that frameworks such as the Parametrized Post Newtonian~\cite{Will:1971zzb, Will:1972zz} are crucial to carry out a comprehensive and conclusive analysis.
It is in this context, that Horndeski gravity became very popular in Cosmology, but let us proceed gradually, starting with the intricate gravitational landscape that emerged after the discovery of cosmic acceleration. 

\subsection{Modifying Gravity}
According to Weinberg-Deser and Lovelock theorems, GR is the unique, Lorentz invariant, low-energy theory of an interacting, massless helicity-2 field in a 4-dimensional space-time.  These theorems help us identify and, to a certain extent, classify the different approaches to beyond GR. The simplest choice is that of introducing an additional scalar field: inspired by inflationary models and the theory of Brans-Dicke~\cite{Brans:1961sx}, the first models to be explored  were extensions of the standard cosmological scenario with the introduction of a dynamical scalar field, \emph{quintessence}, minimally or non-minimally coupled to gravity~\cite{Copeland:2006wr}. There followed  k-essence models, in which the field had a non standard (non-linear) kinetic term to aid tracker solutions that would avoid fine-tuning~\cite{Armendariz-Picon:2000nqq,Armendariz-Picon:2000ulo}.  On the MG side,  models were built via the inclusion of non-linear functions of \emph{Lovelock scalar invariants} \footnote{These scalars are special combinations and contractions of the Riemann tensor which, if present in the Lagrangian,  only introduce second-order derivative contributions to the equations of motion.}~\cite{Lovelock:1971yv} in the action for gravity, the most famous example being $f(R)$ gravity~\cite{DeFelice:2010aj}. A  Lagrangian written in terms of linear and non-linear functions of these scalars propagate only additional scalar degrees of freedom (DOFs), and does not excite extra, ghost-like spin-2 fields. This property makes models with Lovelock scalars interesting candidate models for inflation  and cosmic acceleration. The addition of a non-linear function of the Ricci scalar, $f(R)$, frees up one scalar DOF, the addition of $f(R,\mathcal{G})$, with $\mathcal{G}$ the Gauss-Bonnet term, frees up two additional scalar DOFs~\cite{DeFelice:2010nf}, and so on.

The theories that we mentioned thus far, are very representative of the models explored in the context of cosmic acceleration, but they certainly do not exhaust the  ways in which one can go beyond the assumptions at the basis of the Weinberg-Deser/Lovelock theorems. One can, for instance,  work in higher-dimensional space-times, give mass to the graviton or break Lorentz-invariance, as thoroughly reviewed in~\citep{Clifton:2011jh}. Here we shall mention one of the earliest braneworld theories considered, the five-dimensional Dvali-Gabadadze-Porrati (DGP) model~\cite{Dvali:2000hr}, where all matter belong to a four-dimensional brane embedded in a 5-dimensional bulk where gravity lives. We shall return to this theory shortly.

While models were being formulated, the community started to explore their predictions not only for the expansion history of the Universe, but also for the dynamics of scalar cosmological perturbations that bring to the formation of LSS.  As a result, we gained  key insights into the prospect of testing gravity with LSS, identifying the optimal  probe combinations to use towards this goal~\cite{Zhao:2008bn,Amendola:2012ky,Sapone:2009mb,Harnois_D_raps_2015,Alonso:2016suf,Leonard:2015hha,Baker:2011jy}. f(R) gravity and the DGP model were among the first to be thoroughly explored~\cite{Song:2006ej,Pogosian:2007sw,Hu:2007nk,Charmousis:2006pn,Song:2007wd,Cardoso:2007xc,Lombriser:2009xg,Seahra:2010fj,Xu:2013ega,Schmidt:2009sv,Bag:2018jle,Schmidt:2009yj}. It quickly emerged that the cosmological phenomenology of DE/MG models is significantly richer than the one in the standard cosmological scenario, $\Lambda$CDM (which relies on GR to describe gravity, and includes  Cold Dark Matter  and the cosmological constant $\Lambda$ as ingredients, along with ordinary matter and radiation). In the latter case, the expansion at late times is driven by the cosmological constant, with equation of state $w=-1$; once inside the cosmological horizon, inhomogeneities in the density of matter grow at the same rate on all scales; correspondingly, structure grows with a rate set uniquely by the Hubble expansion; weak gravitational lensing (i.e. the bending of light from background galaxies by foreground LSS) proceeds in a way consistent with clustering, with relativistic and non-relativistic particles following the same geodesics. These  relations are generally broken in models of DE and MG: the dynamics of the background can be characterized by a $w\neq -1$, and more generally an evolving $w(z)$; the  growth of structure is not uniquely set by the expansion rate, and can proceed in a scale-dependent way; lensing and clustering are not necessarily equivalent. These are all consequences of the fact that a new degree of freedom, typically a scalar field, has been introduced (or activated) in the theory; this field generally introduces a characteristic length scale and mediates an extra force, \emph{fifth force}, between matter particles.

These realizations, gradually inspired the formulation of a phenomenological framework for cosmological tests of gravity in terms of the following functions: the already popular equation of state of DE, $w(z)$, to parametrize any deviation at the level of background dynamics; along with $\mu(k,z)$ and $\Sigma(k,z)$ \footnote{Here $z$ is the redshift, and $k$ represents the  wavenumber associated to a given scale in Fourier space.}, functions of time and scale, to capture deviations from $\Lambda$CDM in the dynamics of cosmological perturbations. Specifically, $\mu$ encodes modifications in the clustering, while $\Sigma$ encodes modifications in the lensing, and together with $w(z)$ they allow to define a complete set of equations to study the dynamics of scalar perturbations on linear scales~\cite{Amendola:2007rr,Bertschinger:2008zb, Pogosian:2010tj}. Anything that observations can tell us about the growth of structure can be stored into $\mu$ and $\Sigma$, and indeed this framework has been widely adopted by different collaborations, e.g. Planck~\cite{Planck:2015bue} and the Dark Energy Survey~\cite{DES:2022ccp}, to place constraints on departures from $\Lambda$CDM with data from the cosmic microwave background and LSS, respectively.
 Yet, there are certain shortcomings: it is very agnostic about the underlying theory (e.g. there is no guarantee that any functional form of these functions will correspond to a viable theory~\cite{Peirone:2017lgi}) and it is restricted to scalar perturbations,  thus overlooking the dynamics of tensor perturbations and missing out on the complementary information contributed by gravitational waves.  These limitations can be overcome if we use  Horndeski gravity to inform theoretically $w(z),\mu(k,z)$ and $\Sigma(k,z)$, as we shall discuss later. Alternative frameworks, such as the Parametrized Post-Friedmann one ~\cite{Baker:2011jy,Baker:2012zs}, have also been explored. 

Meanwhile, the exploration of theoretical scenarios continued. Studying the decoupling limit of the braneworld DGP model, Nicolis et al.~\cite{Nicolis:2008in} realized that the scalar field corresponding to the bending mode of the brane obeyed the Galilean shift symmetry, inherited from the invariance under  higher-dimensional Lorentz transformations: $\partial_\mu\phi\rightarrow\partial_\mu\phi+b_\mu$, where $b_\mu$ are constants. The corresponding Lagrangian contained  derivative coupling terms of cubic order in the scalar field, e.g. $\Box\phi\,\partial\phi\cdot\partial\phi$, but still  led to second-order equations of motion. Models of massive gravity in four dimensions lead to analogous terms in the decoupling limit,  but at quartic and quintic order in the scalar field; see the dRGT theory~\cite{deRham:2010kj} for a ghost-free massive gravity model which produces such terms.
Interestingly, requiring a theory for a scalar field to be Galilean invariant and to have equations of motions at most of
2nd order, identifies a finite number of terms that can enter the corresponding Lagrangian. Very much like for Lovelock gravity!
In this case, we have $D+1$ Galileon terms for a Lagrangian in $D$ dimensions, following the general structure $\partial\phi\cdot\partial\phi(\partial^2\phi)^{n-2}$ with $n=1,\dots,D+1$. Expressions of order $n>D+1$, while still being Galilean invariant, correspond to surface terms that do not contribute to the equations of motion in $D$ dimensions.
All this held in flat, Minkowski space. If we extend the analysis to  curved space-time, the Lagrangians need to be promoted to a covariant form. In 2009, Deffayet et al.~\cite{Deffayet:2009wt} derived the corresponding covariant Lagrangians that keep the field equations of second-order, while generally breaking the Galilean symmetry (albeit, in most cases, softly). The result is the Covariant Galileon action:
\begin{eqnarray}
    \label{cov_galileon}
    S_{CG}&=&\int{}d^4x\sqrt{-g}\left\{\frac{M_P^2}{2}R-\frac{1}{2}c_1M^3\phi+c_2 X+\frac{2c_3}{M^3} X \Box\phi+ \frac{c_4}{M^6} X^2 R +\frac{2c_4}{ M^6}X\left[ (\Box\phi)^2 - \phi^{\mu \nu} \phi_{\mu \nu}\right]\right.\nonumber\\
&&\left.-\frac{3c_5}{M^9} X^2 G_{\mu \nu}\phi^{\mu \nu}+\frac{c_5}{M^9}X\left[ (\Box\phi)^3 -3\Box\phi \,  \phi^{\mu \nu} \phi_{\mu \nu}+2   \phi^{\mu \nu} \phi_{\mu \sigma}\phi_{\sigma}^{\nu}\right]\right\}\,,
\end{eqnarray}
where $M_P^2$ is the Planck mass, $X$ is the kinetic term of the scalar field, $X\equiv -\frac{1}{2}\partial_\mu\phi\partial^\mu\phi$,  greek subscripts and superscripts indicate covariant derivatives, the mass scale $M$ is related to the horizon scale by $M^3=M_PH_0^2$ and  $c_2,c_3,c_4,c_5$ are dimensionless constants. We can use the latter to identify 5 blocks inside the Lagrangian, each one separately contributing second-order equations of motion. In particular, the non-minimal couplings, $X^2R$ and $X^2G_{\mu\nu}\phi^{\mu\nu}$,  and the specific linear combinations of operators in the $c_4$ and $c_5$ terms, respectively, are necessary in order to avoid higher derivatives in the equations of motion.

With the breaking of the Galilean symmetry, more freedom is available in setting the coefficients of each $n$-th order term in the Lagrangian, while still maintaining second-order equations of motion. The most general scalar-tensor field theory with an action that depends on derivatives of order two or less of the fields, and results in second-order equations of motion in a general curved space-time was identified in~\cite{Deffayet:2011gz} and is known as Generalized Galileons:
\begin{equation}
    \label{Gen_Galileon_Lagrangian}
    S_{\rm GG} = \int d^4x \sqrt{-g} \sum_{i=2}^{5}{\cal L}_{i} \ ,%+ {\cal L}_M(g_{\mu \nu},\psi)
\end{equation}
with
\begin{eqnarray}
    \label{Gen_Galileons}
{\cal L}_{2} & = & G_2(\phi,X),\nonumber\\ 
{\cal L}_{3} & = & -G_{3}(\phi,X)\Box\phi,\nonumber\\ 
{\cal L}_{4} & = & G_{4}(\phi,X)\, R+G_{4X}\,[(\Box\phi)^{2}-(\nabla_{\mu}\nabla_{\nu}\phi)\,(\nabla^{\mu}\nabla^{\nu}\phi)]\,,\nonumber\\ 
{\cal L}_{5} & = & G_{5}(\phi,X)\, G_{\mu\nu}\,(\nabla^{\mu}\nabla^{\nu}\phi)-\frac{1}{6}\, G_{5X}\,[(\Box\phi)^{3}-3(\Box\phi)\,(\nabla_{\mu}\nabla_{\nu}\phi)\,(\nabla^{\mu}\nabla^{\nu}\phi)\nonumber \\
&+&2(\nabla^{\mu}\nabla_{\alpha}\phi)\,(\nabla^{\alpha}\nabla_{\beta}\phi)\,(\nabla^{\beta}\nabla_{\mu}\phi)] \ ,
\end{eqnarray}
where $G_i(\phi,X)$ are free functions of the scalar field and its kinetic term, and  $G_{iX}=\partial G_i/\partial X$. Theories are referred to as cubic, quartic and quintic depending on whether $G_3,G_4,G_5$ are included, respectively.  Models for which $\partial G_i/\partial\phi=0$ are commonly referred to as shift-symmetric,  due to their invariance under $\phi\rightarrow\phi+c$.

Galileon models gained significant interest in Cosmology because they allow for self accelerating solutions that could describe both the inflationary epoch and the late time accelerated expansion~\cite{Kobayashi:2011nu,DeFelice:2010nf,Chow:2009fm}. In fact, action~(\ref{Gen_Galileons}) includes as subcases most of the models previously explored in the context of inflation and cosmic acceleration, including those mentioned at the beginning of this subsection. It reduces to the Covariant Galileons when $G_2=-\frac{c_1}{2}M^3\phi+c_2X, G_3=-\frac{2c_3}{M^3}X, G_4=\frac{M_P^2}{2}+\frac{c_4}{M^6}X^2, G_5=-\frac{3c_5}{M^9}X^2$.

Galileons have the interesting property that the $(n+1)$-th galileon Lagrangian corresponds to $(\partial\phi)^2$ times the equation of motion of the $n$-th galileon Lagrangian. This property is referred to as \emph{Euler hierarchy}~\cite{Fairlie:1992nb,Fairlie:1991qe,Fairlie:1992he}. Another property of Generalized Galileons is non-renormalization, i.e. the fact that they do not receive quantum corrections at any order in perturbation theory~\cite{Luty:2003vm,Hinterbichler:2010xn}. 

\subsection{The re-emergence of Horndeski Gravity}
\begin{figure}
\centering
\includegraphics[width=0.6\textwidth]{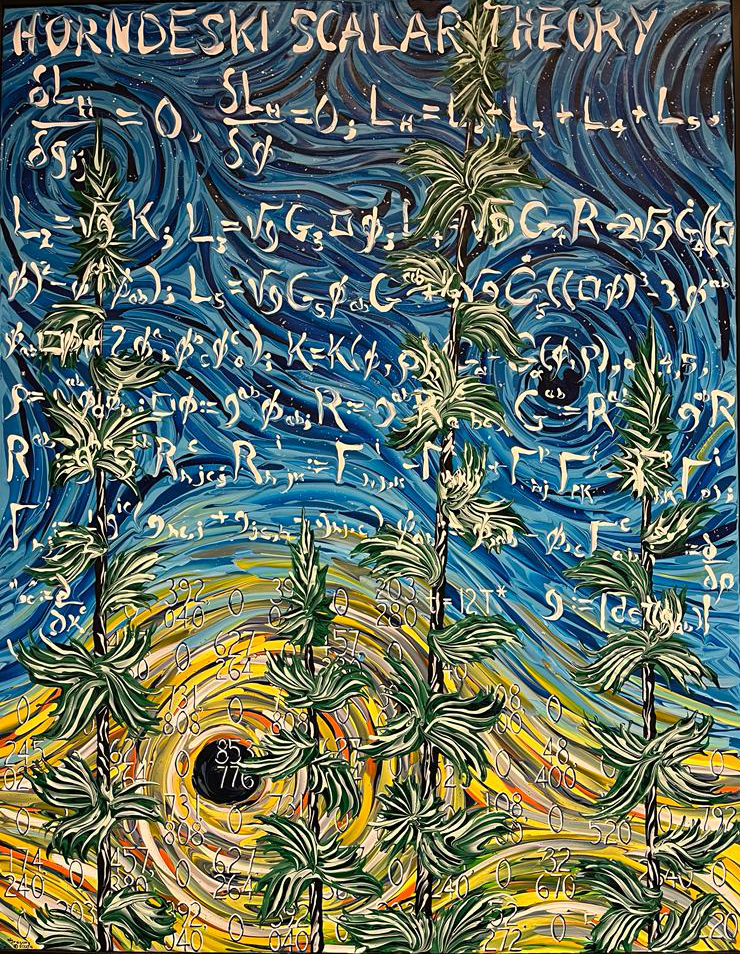}
 \caption{"Horndeski Scalar Theory, Past, Present and Future." (2016),  painting by Gregory Horndeski depicting the equations of Horndeski gravity.}
\label{fig:horndeski_gravity}
\end{figure}

In 2011 Charmousis et al.~\cite{Charmousis:2011bf} revisited Horndeski gravity, with the goal of studying it on a FLRW background and looking for subclasses of it that would incorporate a viable self-tuning mechanism as an approach to the cosmological constant problem. Their paper brought Horndeski gravity back into light, making it known to the cosmological community. It also highlighted how most of the candidate scalar-tensor models of DE and MG were special cases of it. It was at that point that  Kobayashi and collaborators, who had just studied the cosmology of Generalized Galileons in~\citep{Kobayashi:2011nu}, noticed the equivalence between  the latter, as formulated in~\cite{Deffayet:2011gz}, and Horndeski gravity. They added a proof of this equivalence in an Appendix  to their original 2011 paper~\citep{Kobayashi:2011nu}, where they showed that upon identifying:
\begin{eqnarray}
    G_2&=& \kappa_9+4X\int^X dX^\prime(\kappa_{8\phi}-2\kappa_{3\phi\phi}),\nonumber\\
    G_3&=&6F_\phi-2X\kappa_8-8X\kappa_{3\phi} +2\int^XdX^\prime(\kappa_{8}-2\kappa_{3\phi})\nonumber\\
    G_4&=&2F--4 X\kappa_3 \nonumber\\
    G_5&=&-4\kappa_1 \nonumber\\
\end{eqnarray}
and performing integration by parts, action~(\ref{Gen_Galileon_Lagrangian}) becomes exactly the original Horndeski action of~(\ref{Horndeski_original}). A painting by Gregory Horndeski, depicting the equations of Horndeski gravity is presented in Fig.~\ref{fig:horndeski_gravity}. More will be said about Fig. 1 below. Note that in this figure $\rho=-2X$. 

\subsection{Screening mechanisms}
Local tests of gravity have provided a spectacular confirmation of GR in the Solar System and around compact objects. Any alternative theory that introduces additional  DOFs must incorporate a \emph{screening mechanism} to  'hide' them in dense regions, effectively reducing to GR. There is only a handful of screening mechanisms and they all rely on the scalar field (i.e. the extra DOF in  most DE/MG models, and certainly in those relevant for this review) developing non-linearities in certain environments. We can organize them in three broad classes: those  which become active in regions of high Newtonian potential, known as Chameleon and symmetron, typical of Generalized Brans-Dicke models~\cite{Khoury:2003aq}; mechanisms in which first derivatives of the field become important in dense regions, typical of k-mouflage models~\cite{Babichev_2009}; and the Vainshtein screening~\cite{Vainshtein:1972sx}, where second derivatives of the field become dominant in dense regions, hallmark of the pure Galileon models. Horndeski gravity contains all these three types of mechanisms: non-minimal couplings like the one introduced by a $G_4(\phi)$, source the Chameleon; non-standard kinetic terms that still involve only first derivatives of the field, e.g. non-linear choices for $G_2(X)$, source the kinetic screening; higher-derivative couplings that enter the cubic, quartic and quintic Lagrangians are at the heart of the Vainshtein mechanism. Which mechanism will be at play for a given Horndeski model, depends on the corresponding  $G_i$ functions, and potentially also on the environment, as discussed in~\cite{Gratia:2016tgq}.

Screening mechanisms produce a rich phenomenology, such as macroscopic violations of the equivalence principle~\cite{Hui:2009kc,Sakstein:2017bws,Bartlett:2020tjd}, modifications  to stellar structure~\cite{Chang:2010xh} and pulsations~\cite{Sakstein:2016lyj} and to the galaxy morphology~\cite{Vikram:2014uza, Desmond:2020gzn}, and much more. For a complete review of the extensive literature on this, we refer the reader to~\cite{Jain:2013wgs, Brax:2021zjq, Sakstein:2020gth}.  Screening is crucial to model when studying growth of structure on smaller, non-linear scales, since it can modify the collapse mechanism and hence the halo formation model. Interestingly, it can leave characteristic imprints on the position of the splashback radius in clusters~\cite{Adhikari:2018izo,Contigiani:2018hbn}. Most studies to date have focused on screening around spherically symmetric, static (or slowly rotating) sources, however it  is known that the efficiency of the mechanism depends on the symmetry of the source. This feature has been exploited towards powerful laboratory tests of Chameleon models~\cite{Brax:2011sv, Burrage:2019yle}. One shall also expect to see some impact of this when upgrading the modeling of the collapse in structure formation from spherical to ellipsoidal. Finally, screening is expected to produce several interesting features around sources which have some dynamics, e.g. scalar gravitational radiation~\cite{Dar:2018dra,Silvestri:2011ch}

\subsection{Gravitational Waves propagation in a Horndeski Universe}
When modifying the action for gravity in the cosmological context, one needs also to take into account possible effects on the propagation of gravitational waves (GWs). Even more so now, that we are witnessing the dawn of the GW Astronomy era, started with the historical first direct detection of a GW from a binary black hole merger in 2015~\cite{Abbott:2016blz}. Generalized Galileon models can impact the propagation of GWs in two ways: modifying their speed and contributing an additional friction term in their propagation equation. The latter further damps the amplitude of the GW, thus affecting the luminosity distance inferred from the GW amplitude, as follows:
\begin{equation}
\label{GW_lum_distance}
\frac{d_L^{\rm GW}(z)}{d_L^{\rm EM}(z)}=\frac{M_*(0)}{M_*(z)}\,,\,\,\,{\rm where}\,\,\,\,\,M^2_*(z)\equiv 2\left[G_4-2XG_{4X}+XG_{5\phi}-\dot{\phi}HXG_{5X}\right]
\end{equation}
where $M_*^2$ corresponds to the running Planck mass of Generalized Galileons, and is induced by the non-minimal coupling terms in the theory; $d_L^{\rm EM}$ is the luminosity distance that would be associated to an electromagnetic source at the same redshift of the GW. This opens the way to interesting  tests of gravity with standard sirens~\cite{Amendola:2017ovw,Belgacem:2017ihm,LISACosmologyWorkingGroup:2019mwx}.

The propagation speed instead receives the following contributions: 
\begin{equation}
\label{cT_constraint}
c^2_T=1+\frac{X\left[2G_{4X}-2G_{5\phi}-\left(\ddot{\phi}-H\dot{\phi}\right)G_{5X}\right]}{G_4-2XG_{4X}+XG_{5\phi}-\dot{\phi}HXG_{5X}}\,.
\end{equation}
These originate from the non-linearities generated by higher order kinetic energy terms, which modify the light-cone structure of GWs with respect to that of photons. Simpler forms of non-minimal couplings, corresponding to $G_4(\phi)$ and $G_5(\phi)$, do not affect the light-cone but contribute to the damping term~(\ref{GW_lum_distance}). Interestingly, all these  components, which do not exist for canonical scalar fields,  generate a non-zero shear component of the scalar field energy-momentum,  which  manifest as an effective anisotropic stress in LSS, introducing a difference between lensing and clustering. For these reasons, the latter is often considered a smoking gun signature of MG since it is in close correspondence with a modified propagation of tensor modes~\cite{Sawicki:2016klv,Matos:2022uew}.

2017 saw the first detection of a GW from a binary neutron star inspiral, GW170817~\cite{TheLIGOScientific:2017qsa} together with its electromagnetic counterpart, GRB170817A~\cite{Savchenko:2017ffs,LIGOScientific:2017ync,Goldstein:2017mmi}. This event allowed to place severe constraints on the speed of tensors, setting it to be equal to that of light~\cite{LIGOScientific:2017zic}. 
As it was pointed out in a series of papers~\cite{Baker:2017hug,Creminelli:2017sry,Ezquiaga:2017ekz,Langlois:2017dyl}, this constraint would effectively rule out a significant portion of the Generalized Galileons, reducing the quartic lagrangian to $G_4(\phi)R$ and eliminating the quintic one. However, it is important to keep in mind that the source associated to this event  is very close by in cosmological terms, at a redshift of  $z\sim0.01$, while cosmological data comes from higher redshifts; and,  as pointed out in~\cite{deRham:2018red}, its detection corresponds to energy scales that are close to the cut-off scale at which low-energy actions like~(\ref{Gen_Galileon_Lagrangian}) become invalid. With all due caveats, the first bright siren, GW170817, certainly delivered a strong message on the importance of the synergy between LSS and GWs when exploring gravity in the cosmological context.

\subsection{Cosmological tests of gravity}
Since its re-emergence in Cosmology, Horndeski gravity  quickly became a popular framework for cosmological tests of gravity and within a few years, the word reached the original author, Gregory Horndeski, as he discusses in the Conclusions of this essay. Shortly after, he created a  painted depiction of the equations of Horndeski gravity, Fig.~\ref{fig:horndeski_gravity}, which was eventually purchased by the Institute Lorentz for Theoretical Physics of Leiden University. As shall be clear from the discussion in the previous subsections, Horndeski gravity encompasses most of the candidate models of DE/MG explored in the decades since the discovery of cosmic acceleration, providing a compact way of organizing them within a unified action. This has the important advantage of facilitating the theoretical and observational comparison of the different scenarios.

What provided an additional boost, was the formulation of the effective field theory of DE (EFTofDE)~\cite{Bloomfield:2012ff, Gubitosi:2012hu,Piazza:2013coa,Frusciante:2019xia}, inspired by the EFT of inflation~\cite{Cheung:2007st} and of quintessence~\cite{Creminelli:2008wc}. EFTofDE builds on fundamental symmetry arguments, treating cosmological perturbations as the Nambu-Goldstone modes  associated to a cosmological state that spontaneously breaks time-translations, as is the case with any model that sources cosmic acceleration in a dynamical way. The formalism identifies a set of Lagrangian operators containing an increasing number of cosmological perturbations and derivatives acting on them.  In its original formulation the EFTofDE identifies the most general Lagrangian, quadratic in perturbations around the Friedmann-Lemaitre-Robertson-Walker (FLRW) Universe, which propagates a massless tensor and a scalar field with second-order equations of motion. In other words,  the quadratic action of Horndeski gravity! Thus providing Horndeski gravity with a new interpretation: 
the low-energy effective action of gravity on the large scales that characterize the expanding FLRW Universe, further confirming that it represents an ideal framework for exploring gravity with LSS!

The EFTofDE action, in conformal time,  is written in terms of a handful of operators:
\begin{eqnarray}
\mathcal{S}_{\rm DE} =&& \int d^4x \sqrt{-g}  \bigg\{ \frac{M_P^2}{2} (1+\Omega(\tau))R + \Lambda(\tau) - c(\tau)\,a^2\delta g^{00} + \frac{M_2^4(\tau)}{2} \left(a^2 \delta g^{00} \right)^2\nonumber\\
 -&& \frac{\bar{M}_1^3(\tau)}{2} a^2\delta g^{00}\,\delta {K}{^\mu_\mu} -  \frac{\bar{M}_2^2(\tau)}{2}\left[\left( \delta {K}{^\mu_\mu}\right)^2 - \delta {K}{^\mu_\nu}\,\delta {K}{^\nu_\mu}-\frac{ a^2}{2} \delta g^{00}\,\delta R^{(3)}\right] \bigg\} \,,
\label{EFT_action}
\end{eqnarray}
where $\Omega, \Lambda, c,M_2^4, \bar{M}_1^3, \bar{M}_2^2$ are free functions of time; we have chosen a specific foliation of space-time by identifying constant-time hypersurfaces with uniform field ones; $K^\mu_\nu$ and $R^{(3)}$ are, respectively,  the extrinsic curvature and  three dimensional spatial Ricci scalar of these  hypersurfaces. As with all previous actions, we are focusing on gravity and the scalar DOF, with the understanding that matter fields, $\chi_i$  appear in an accompanying action of the form $S_m=\int d^4x\sqrt{-g}\mathcal{L}_m[g_{\mu\nu},\chi_i]$.
Following the EFT prescription, one could include further operators in the expansion in field and derivatives, extending to Beyond Horndeski. In fact, there is a handful of operators that one can add to the action while maintaining the resulting equations of motion of second-order for the propagating DOFs, thanks to degeneracy properties of the kinetic matrix as it was  shown in~\citep{Gleyzes:2014dya,Gleyzes:2014qga}  where the authors identified the so-called GLPV theories. Healthy Beyond Horndeski theories can also be identified by applying a disformal transformation~\cite{Bekenstein:1992pj} to the Horndeski Lagrangian~\cite{Zumalacarregui:2013pma}. The broad class of Degenerate Higher Order Scalar-Tensor theories, that encompasses also the former cases, was introduced in~\cite{Langlois:2015cwa,BenAchour:2016cay,Motohashi:2016ftl}. See~\cite{Traykova:2019oyx,Hiramatsu:2022fgn,Sugiyama:2023tes} for some recent explorations of the cosmological phenomenology of these models.  For a more exhaustive overview of Beyond Horndeski models, we refer the reader to~\cite{Langlois:2018dxi}, and references therein. 
Interestingly, some of the Beyond Horndeski terms in the action can be constrained by considering the potential decay of GWs into DE, a first discussed in~\cite{Creminelli:2018xsv,Creminelli:2019kjy}. 

Equation~(\ref{EFT_action}) is equivalent to the Horndeski action expanded to quadratic order in perturbations, and in fact any theory belonging to the Horndeski family, can be uniquely mapped into~(\ref{EFT_action}) via a simple procedure, as reviewed in~\citep{Frusciante:2019xia}. We refer the reader to the latter reference for the complete mapping prescription, and report here only the expressions for some of the EFT functions that will come in handy for the discussion below:
\begin{equation}
    \label{GG_mapping}
   M_P^2\left(1+\Omega\right)=M_*^2-\bar{M}_2^2=2\left[G_4-XG_{5\phi}-X\ddot{\phi}G_{5X}\right]\,,\,\,\,\,\,\,\bar{M}_2^2= -2X\left[2G_{4X}-2G_{5\phi}-\left(\ddot{\phi}-H\dot{\phi}\right)G_{5X}\right]\,.
\end{equation}

Thanks to the mapping, Eq.~(\ref{EFT_action}) provides a unifying action to explore at once the cosmological dynamics of a broad landscape of DE/MG models; one can work both in a model specific way, resorting to the mapping procedure, or adopt a more agnostic approach, exploring the phenomenology produced by different choices of five free functions of time $\Omega, \Lambda, M_2^4, \bar{M}_1^3, \bar{M}_2^2$ (the Friedmann equation provides a constraint equation for the function $c$); see~\cite{Raveri:2014cka,Bellini:2015xja,Raveri:2019mxg,Frusciante:2018jzw,Melville:2019wyy,SpurioMancini:2019rxy,Noller:2018wyv,Kreisch:2017uet,Noller:2020afd} for some examples of constraints with current data. In terms of (numerical) feasibility, the latter represents a significant improvement with respect to using the generic functions $G_i(\phi,X)$. These characteristics were exploited in the creation of \texttt{EFTCAMB}~\cite{Raveri:2014cka,Hu:2013twa}, a patch that implements EFTofDE into the public Einstein-Boltzmann solver \texttt{CAMB}~\cite{Lewis:1999bs}. This can be used to create precise predictions for the phenomenology of LSS and GWs on linear, cosmological scales for any Horndeski gravity model which can then be translated into corresponding functional forms of $w(z), \mu(k,z), \Sigma(k,z)$~\cite{Peirone:2017ywi, Perenon:2016blf}. 

Furthermore, action~(\ref{EFT_action}) facilitates the simultaneous analysis of  LSS and the propagation of GWs on the cosmological background, which is a crucial improvement given the recent advent of GW Cosmology. It allows us to factor any constraints from the direct detection of GWs, like those discussed in the previous subsection, into the exploration of LSS phenomenology. Using the mapping prescription~(\ref{GG_mapping}), the constraint on the speed of GWs, translates very simply into $\bar{M}^2_2=0$, correspondingly simplifying the EFTofDE action; $\Omega$ is connected both to the running Planck mass, $M_*^2$, and to $\bar{M}_2^2$. In the case of luminally propagating tensors, $\Omega$ reduces to $M_*^2$, corresponding to the additional friction term that further damps the amplitude of GWs. As shown in~\cite{Peirone:2017ywi}, the $c_T^2=1$ constraint can have a significant impact on the LSS phenomenology, limiting the possibility for a difference between $\mu$ and $\Sigma$, i.e. between clustering and lensing, to the smaller scales only~\cite{Pogosian:2016pwr}. Another opportunity offered by~(\ref{EFT_action}), is that of studying the joint power of galaxy surveys and future GW surveys, such as LISA~\cite{LISA:2017arv} and Einstein Telescope~\cite{Maggiore:2019uih}, in constraining gravity~\cite{Balaudo:2023klo,Balaudo:2022znx}. 

Starting from a unifying action like~(\ref{EFT_action}),  allows also to perform a very efficient stability analysis, by further expanding the different operators to quadratic order in the perturbations to the metric, and looking at the properties of the resulting kinetic, gradient and mass matrices for the propagating DOFs. This leads to a set of inequalities involving the  free functions in the action and their time derivatives, that represent general criteria of viability~\cite{Gleyzes:2013ooa,Frusciante:2016xoj}.  The latter can provide  additional  constraining power when exploring gravity with cosmological data~\cite{Raveri:2014cka, Salvatelli:2016mgy,Melville:2019wyy} and, importantly, physically inform the  parametrizations of the phenomenological functions, thus ensuring that we sample only regions of the parameter space that correspond to viable theories~\cite{Peirone:2017lgi}. In fact, the combination of Horndeski gravity, written in terms of EFTofDE,  with the phenomenological framework $(w,\mu,\Sigma)$,  crucially addresses the above mentioned shortcomings of the latter, and  can provide insights into what can be learned about gravity from cosmology. In~\cite{Pogosian:2016pwr} for instance, a \emph{Horndeski conjecture} was identified for $\mu$ and $\Sigma$ through an analytical exploration of the implications of  Horndeski gravity for these two functions. More interestingly, action~(\ref{EFT_action}), in combination with \texttt{EFTCAMB},  allows a Monte Carlo sampling of the viable gravitational landscape, leading to the formulation of powerful \emph{theoretical priors} for the phenomenology of LSS~\cite{Espejo:2018hxa, Peirone:2017ywi,Raveri:2017qvt,Traykova:2021hbr}; priors that can also take into account constraints on the speed of tensors from the direct detection of gravitational waves. All these aspects were used in a recent reconstruction of gravity from currently available cosmological data~\cite{Pogosian:2021mcs,Raveri:2021dbu}, which besides showcasing the potentiality of current data, sets the stage for future tests with LSS surveys.

Much more could be said, but it would elude the purpose of this short, celebratory essay. We refer the reader to~\cite{Kobayashi:2019hrl} for a more technical review of Horndeski and Beyond Horndeski. We will conclude this part, noting how  50 years after its formulation, Horndeski theory  has gained a central role in cosmological tests of gravity. 
Exploring Horndeski models has been insightful so far, but the true potential of this framework will be at display with Stage IV LSS surveys that are seeing light in these years. 

\section{Concluding Remarks}
I "retired" from Physics to pursue a career as an artist in  1981, maintaining only a passing interest in the developments of General Relativity.  Then in July, 2014 I received a phone call from Prof. Jim Isenberg, who was a Post-Doc at the University of Waterloo in the late 1970's.  Jim and I did some work on my vector-tensor theory of gravity and electromagnetism~\cite{Horndeski:1976gi}. Jim told me that he had just gotten back from a relativity conference in Italy, and while walking down a hallway during the conference he had overheard some physicists discussing Horndeski theory. So he interrupted their conversation and asked if they were talking about Horndeski's vector theory.  They replied no, they were interested in his scalar theory.  Jim was bowled over by that, since he never knew that I did anything on scalar-tensor theory. He suggested that I Google Horndeski scalar theory to find out what was going on.  I did and was amazed to find out that there were 242 citations to my paper! Then about a year later I heard from a young physicist, Dr. Alejandro Guarnizo-Trilleras, who wanted to use the image of one of my paintings on the back cover of his thesis. I, of course let him do that, and then in the course of reading his thesis I discovered that some physicists were "going beyond Horndeski Theory"~\cite{Langlois:2015cwa,Gleyzes:2014dya,BenAchour:2016fzp}.  Well, as long as I am alive no one is going beyond Horndeski Theory without me.  That is when I got back into doing research on scalar-tensor theory. For an excellent review of Beyond Horndeski theory please see~\cite{Kobayashi:2019hrl}.   Rather than remark on all the various things I have been exploring involving scalar-tensor theories since 2015, I would like to mention that I have  been working on the bi-scalar tensor generalization of "conventional" Horndeski theory. A great deal of progress on this problem has been made by Ohashi, $et$ $al,$in~\cite{Ohashi:2015fma}.  They have found the basic form of $A^{\alpha\beta}$ for second-order bi-scalar tensor field theories in a four-dimensional space, but were unable to find a Lagrangian that could generate $A^{\alpha\beta}$.  I feel fairly comfortable that I have found that Lagrangian and it will appear in a future publication.  Hopefully we won't have to wait another 30 years before physicists find a use for that result.

The work presented in the previous section by Prof. Silvestri clearly illustrates that the equations I developed back in the 1970s have proved to be immensely valuable to various lines of research in Cosmology. One of the main reasons for their usefulness has been the versatility they offer due to the presence of the four coefficient functions of $\phi$ and X appearing in them.  So what I originally surmised to be the main drawback of my scalar-tensor equations, has actually turned out to be their saving grace.

\acknowledgments
We are grateful to Justin Khoury and Tsutomu Kobayashi for their precious feedback and input. We would also like to thank Prof. A. Wipf, Editor in Chief of the International Journal of Theoretical Physics, for providing us with the opportunity to present this essay on Horndeski gravity, and also for the patience he has evinced as we prepared it. AS acknowledges  support from the NWO and the Dutch Ministry of Education, Culture and Science (OCW) (through NWO VIDI Grant No. 2019/ENW/00678104 and ENW-XL Grant OCENW.XL21.XL21.025 DUSC) and from the D-ITP consortium.
\\
\\

\bibliography{Horndeski_50}

\end{document}